\documentclass[pra,twocolumn,showpacs,showkeys,linenumbers]{revtex4}

\usepackage{amssymb,amsfonts,amsmath,enumerate}
\usepackage{graphicx}
\newcommand{\tr}{\mathrm{Tr}}

\begin{document}

\title{
On the connection between Complementarity and Uncertainty Principles in the Mach--Zehnder interferometric setting
}

\author{G.M.\ Bosyk, M.\ Portesi, F.\ Holik, A.\ Plastino}
\affiliation{Instituto de F\'{\i}sica La Plata, CONICET, and Departamento de F\'{\i}sica, Facultad de Ciencias Exactas, Universidad Nacional de La Plata, 115 y 49, C.C.~67, 1900 La Plata, Argentina}

\begin{abstract}

We revisit, in the framework of Mach--Zehnder interferometry, the
connection between the complementarity and uncertainty principles
of quantum mechanics. Specifically, we show that, for a pair of
suitably chosen observables, the trade-off relation between the
complementary path information and fringe visibility is equivalent
to the uncertainty relation given by Schr\"odinger and Robertson,
and  to the one provided by Landau and Pollak as well. We also
employ entropic uncertainty relations (based on R\'{e}nyi entropic
measures) and study their meaning for  different values of the
entropic parameter. We show that these different values define
regimes which yield qualitatively different information concerning
the system, in agreement  with  findings of [A.~Luis, Phys.\ Rev.\ A~{\bf
84}, 034101 (2011)]. We find that there exists a regime for which
the entropic uncertinty relations can be used as criteria to
pinpoint  non trivial states of minimum uncertainty.

\pacs{03.65.Ca, 03.65.Ta, 02.50.-r, 05.90.+m}

\keywords{Wave--particle duality, Complementarity, Uncertainty relation}

\end{abstract}

\date{\today}

\maketitle

\section{Introduction}

The Complementarity Principle (CP)~\cite{Bohr1937} lies at the heart
of Quantum Mechanics (QM). Many years have passed since its original
formulation but, still today, there is an important debate regarding
its adequate interpretation and its precise definition in several contexts~\cite{Jaeger1995,Englert1996,Busch2006,Liu2012}.

The Complementarity Principle has been both
theoretically%~\cite{Busch2006} 
and experimentally%~\cite{experim}
studied in the framework of Mach--Zehnder (MZ) interferometrics. The
MZ framework is particularly suitable for discussions regarding
wave--particle duality, and there is a debate concerning the
complementarity of the fringe-visibility observable (wave aspect)
and the which-way-has-passed question (particle aspect).
In this regard, the wave and particle properties are represented by measurable
quantities $P$ and $V$, respectively, which satisfy the duality relation~\cite{Jaeger1995,Englert1996}
\begin{equation}\label{CP}
P^2+V^2 \leq 1 .
\end{equation}
Despite the fact that this quantitative formulation of the
complementarity principle is expressed in a way that resembles
inequalities typical of the uncertainty principle, the derivation
of Eq.~\eqref{CP} does not involve any mention of inherent
fluctuations in the measured quantities. Inspired in this
quantitative similarity, we are interested in looking deeper at
the connection between these two important principles of Quantum
Mechanics. Specifically, we address the question: is
Eq.~\eqref{CP} the expression of an uncertainty relation? This
basic issue has been the subject of intense debate in the
literature. Answers in both the affirmative and the negative have been
provided by various authors~(see, for instance,
Refs.~\cite{Durr2000,Busch2006,Bjork1999,Luis2001}).
Our goal is, with regards to Eq.~\eqref{CP}, to shed some new
light on the issue by considering several ways of quantifying
uncertainty, concentrating attention on variance-based and
entropy-based inequalities.

The article's  outline  is as follows: in
Sect.~\ref{s:Mach-Zehnder} we review details of the discussion
concerning the duality relation~\eqref{CP}, introducing relevant
operators that account for the path information and
fringe visibility in double-slit-like experiments.
Sect.~\ref{s:UncertaintyRelations} is devoted to summarize various
formulations of the uncertainty principle that were applied to our
problem, i.e.\ for a pair of two-level discrete operators, by
employing variances as well as entropic and other measures. In
Sect.~\ref{s:Relationship} we provide an affirmative answer to the
question posed in the case of the uncertainty inequalities
prescribed by Shr\"odinger--Robertson and by Landau--Pollak,
demonstrating the full equivalence between them. Additionally, our
analysis of a class of entropic uncertainty inequalities shows
that they are not on the same footing as the above ones but that
they yield nonetheless nontrivial information about the system.
This is achieved  by studying  states that saturate the
entropic inequality. We find that, according to the value of the
R\'{e}nyi parameter,  different regimes can be discerned, a fact
that can be interpreted as giving support to previous
investigations~\cite{Luis2011}. Finally, some conclusions are
drawn in  Sect.~\ref{s:Conclusions}.

\section{Mach--Zehnder
interferometer scheme and complementarity relation}\label{s:Mach-Zehnder}

\begin{figure}[htbp]
\begin{center}
\includegraphics[width=8cm]{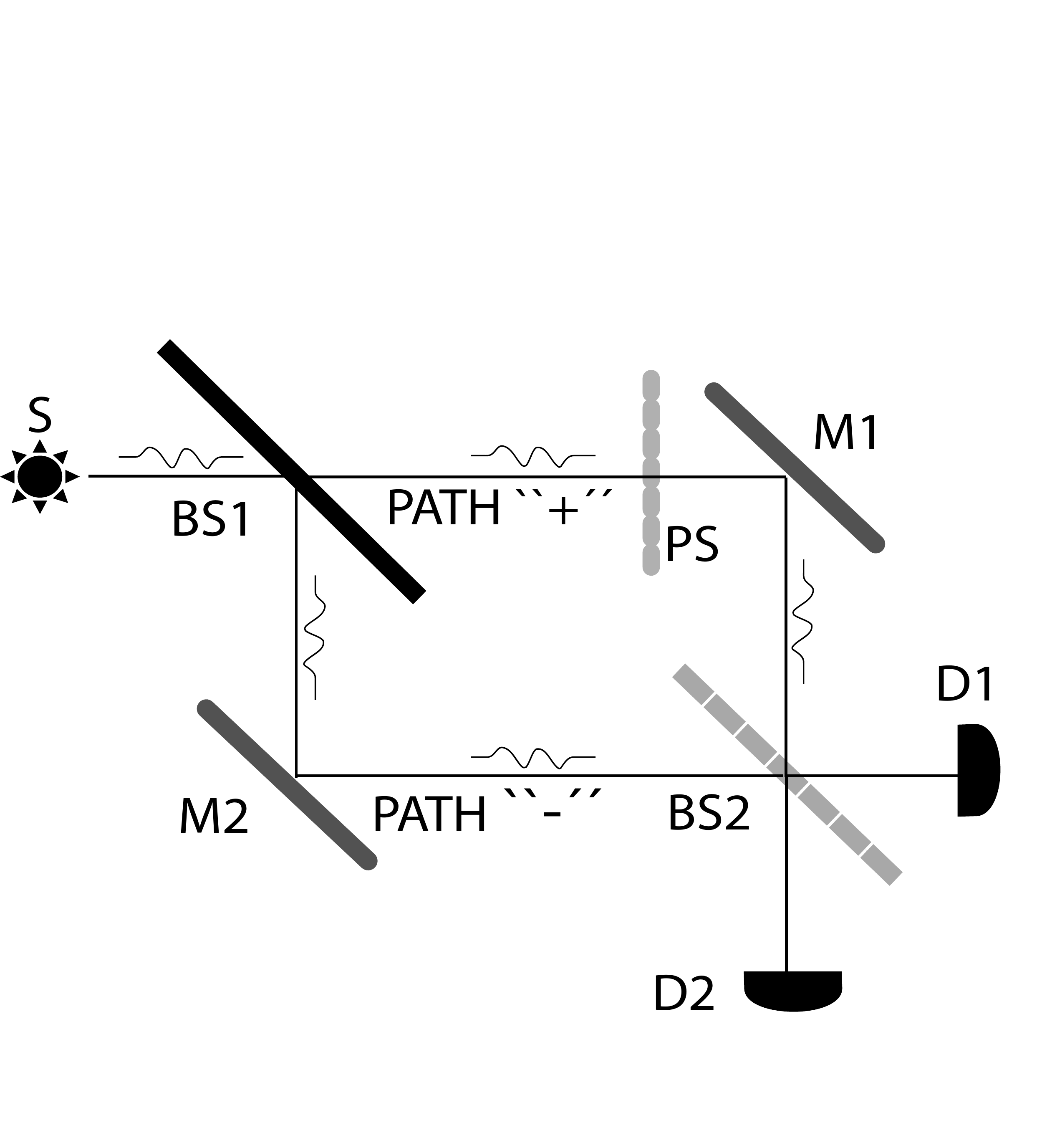}  %%%
\end{center}
\caption{A source S emits a photon which splits, after passing
through the beam splitter BS1, into paths ``+" and ``-". It
reflects in mirrors M1 and M2, and is finally observed using
detectors D1 and D2. A phase
shifter PS and another beam splitter BS2 may be inserted into the
set-up, in order to produce interference.}
\label{f:figure1}
\end{figure}

The Mach--Zehnder interferometer (Fig.~\ref{f:figure1}) is a device
that has been used in several branches of physics, in particular,
for the study of the Complementarity Principle. An important
quantity is the ``\textit{which way}'' information, that is
quantified by the \textit{predictability} $P$ defined as $P=2L-1$,
where $L=\max\{w_{+},w_{-}\}$, and $w_{+}$ and $w_{-}$ are the
probabilities of the particle taking path ``+" or path ``-",
respectively. On the other hand, the \textit{fringe visibility} is
quantified via a natural extension of the usual measure for intensity of light, that is $V=\frac{p_{\max}-p_{\min}}{p_{\max}+p_{\min}} $
where $p$ stands for the probability that the particle be detected
in some position in space, with $p_{\max}$ and $p_{\min}$ denoting,
respectively, the maximum and minimum of this probability.

The quantitative formulation of CP in the MZ-interferometer scheme
is the celebrated duality relation~\cite{Jaeger1995,Englert1996}
given by Eq.~\eqref{CP}, where the equal sign holds (only) for
pure states. This relation was also implicitly alluded to in the
pioneering works of Refs.~\cite{Wootters1979,Mandel1991}.

The MZ interferometer, having two relevant spatial modes, can  be represented by a two-dimensional Hilbert space
spanned for instance by $\{|0\rangle, |1\rangle \}$, which is the so-called
computational basis. The states $|0\rangle$ and $|1\rangle$ are
eigenstates of the Pauli spin operator $\sigma_z$,
representing the two paths. We use the Bloch representation to
describe quantum density operators as
\begin{equation}\label{e:stateBloch}
\rho = \frac {I+ \vec{s} \cdot \vec{\sigma}}{2}
\end{equation}
where $\vec{\sigma} = (\sigma_x,\sigma_y,\sigma_z)$ denote the Pauli
matrices, $I$ is the $2\times 2$ identity matrix, and $\vec{s}=
(s_x,s_y,s_z)$ is the Bloch vector (with $\|\vec{s}\| \leq 1$) that
characterizes the state. The action of the \mbox{50:50} beam
splitter can be described by the unitary transformation
$U_{\mathrm{BS}}=e^{-i \pi \sigma_y/4}$, which implies a rotation of
$\pi/2$ of the Bloch vector around the $y$-axis. The phase shifter
$U_\phi = e^{-i\phi\sigma_z/2}$ introduces a phase difference equal
to $\phi$ between the paths.

Following Ref.~\cite{Bjork1999}, a sharp observable $\hat{P}$ can be
associated to the predictability, while two families of sharp observables $\hat{V}_\phi$ and
$\hat{V}_\phi^\perp$ can be associated to the visibility. It is possible
to express these operators in terms of the Pauli spin ones as
\begin{eqnarray}
\hat{P} &=& \sigma_z \label{Pop}\\
\hat{V}_\phi &=& (\cos\phi) \ \sigma_x + (\sin\phi) \ \sigma_y \label{Vop1}\\
\hat{V}_\phi^{\perp}&=& -(\sin\phi) \ \sigma_x + (\cos\phi) \ \sigma_y \label{Vop2}
\end{eqnarray}
with $\phi$ ranging, in principle, between 0 and $2\pi$.
Note that $\hat{P}$, $\hat{V}_\phi$,
and $\hat{V}_\phi^\perp$ are a set of mutually complementary
observables, that is, if one is certain about the value of one
observable, then maximum ignorance reigns concerning the value of any of the
other two.

For a system in state $\rho$ (with Bloch vector $\vec s$) the predictability $P$
is obtained by taking the modulus of the expectation value of
observable $\hat{P}$: \ $P=|\langle \hat{P}\rangle |= |s_z|$. The
visibility $V$ can be derived either from the observable
$\hat{V}_\phi$ or from $\hat{V}_\phi^\perp$ by properly choosing
the parameter~$\phi$. Defining $\rho$ as
  $\left(
    \begin{array}{cc}
      \omega_+ & r e^{-i\theta} \\
      r e^{i\theta} & \omega_- \\
    \end{array}
  \right)$,
the auxiliary state variables $r\equiv\frac 12\sqrt{s_x^2+s_y^2}$
and $\tan\theta \equiv\frac{s_y}{s_x}$ allow  us to write \ $\langle
\hat{V}_\phi\rangle=2r \cos( \theta - \phi)$ and \ $\langle
\hat{V}_\phi^\perp\rangle=2r \sin( \theta - \phi)$. Thus the
visibility, which is given by the maximum absolute expectation value
of these observables, is equal to $2r$ and can be obtained using
$\hat{V}_\phi$ if one sets $\phi=\theta$, or $\hat{V}_\phi^\perp$
setting $\phi=\theta-\pi/2$ (arranging the apparatus with a phase
difference of $\pi$ with respect to these angles gives also the same
value of visibility). Finally, due to the positivity of the density
matrix, the complementarity relation~\eqref{CP} is directly
obtained:
\begin{equation}
P^2+V^2= s_x^2+s_y^2+s_z^2 \leq 1 ,
\end{equation}
and it is seen to be saturated whenever $\|\vec{s}\|=1$, i.e.\ for any pure state.

We note that the measurements of the two observables~\eqref{Pop}
and~\eqref{Vop1}, or~\eqref{Pop} and~\eqref{Vop2}, can only be
carried out in two {\it incompatible} experimental set-ups and that
joint measurement is not involved. Therefore the trade off relation~\eqref{CP} expresses
the \emph{preparation complementarity}~\cite{Busch2006}, that is the
impossibility to prepare the system in a state where the two
observables have simultaneously sharp values.

\section{Uncertainty relations}\label{s:UncertaintyRelations}

The Uncertainty Principle~(UP) states that the probability
distributions associated to the outcomes of two incompatible observables cannot be
simultaneously sharp. Quantitative formulations of the UP are known
as \emph{uncertainty relations}~(UR), and there is now a collection of
inequalities that express this principle (see, for instance, the
recent review articles~\cite{Wehner2010,Birula2011}). Before
summarizing the UP formulations to be employed, let us introduce the relevant quantities and fix the
notation.

In general, the
state of an $N$-level system is described by a density operator
$\rho$, with $\tr\rho=1$ and $\rho \geq 0$. Physical observables like $A$
and $B$ are represented by Hermitian operators which in their spectral
decomposition can be written as $A = \sum_{i=1}^N a_i
|a_i\rangle \langle a_i| $ and $B = \sum_{i=1}^N b_i |b_i\rangle
\langle b_i|$, where $a_i$ and $b_i$ are real numbers, and
$\{|a_i\rangle\}_{i=1}^N$ and $\{|b_i\rangle\}_{i=1}^N$ are the
corresponding eigenbases. The probability to obtain a certain value
$a_i$ of observable $A$ is given by Born rule:
$p(A=a_i)=\tr(\rho |a_i\rangle \langle a_i|)$. The so-called overlap between operators $A$ and $B$ is
defined by $c=\max_{i,j} |\langle a_i| b_j\rangle|$ and lies between $\frac{1}{\sqrt N}$ and 1.

In the particular case of two-level or qubit systems
($N=2$), one can use the Bloch representation. Hence, the density operator is given by Eq.~\eqref{e:stateBloch}. Similarly, we can write operators $A$ and
$B$ as
\begin{eqnarray}
A &=& \alpha_1 I + \alpha_2 \, \vec a \cdot \vec \sigma \label{A}\\
B &=& \beta_1 I + \beta_2 \, \vec b \cdot \vec \sigma \label{B}
\end{eqnarray}
where $\alpha_i$ and $\beta_i$ are real numbers, and $\vec
a$ and $\vec b$ are unit vectors on the Bloch sphere. Therefore, in
this representation the probability distributions associated to both observables
take the simple form
\begin{eqnarray}
\{p(A)\} &=& \left\{ \frac{1+ \vec a \cdot \vec s}{2}, \frac{1- \vec a \cdot \vec s}{2} \right\} \label{pA} \\
\{p(B)\} &=& \left\{ \frac{1+ \vec b \cdot \vec s}{2}, \frac{1- \vec b \label{pB}
\cdot \vec s}{2} \right\}
\end{eqnarray}
for a qubit characterized by the Bloch vector $\vec s$. Meanwhile, the overlap is
\begin{equation}
c=\frac{1+ | \vec a \cdot \vec b|}{2} \ \in \ \left[\frac{1}{\sqrt2},1\right]
\end{equation}
where the case $c=1/\sqrt 2$ corresponds to $A$ and $B$ being \emph{complementary observables}.

\subsection{Variance-based uncertainty relations}

Heisenberg, in his famous 1927 paper~\cite{Heisenberg1927}, was the
first to propose an uncertainty relation for position and momentum
observables in terms of their variances. The generalization of Heisenberg inequality for any
arbitrary pair of Hermitian operators  $A$ and $B$ is due
to Robertson~\cite{Robertson1929} and contains the commutator $[A,B]$. A further tighter relation was derived by
Schr\"odinger~\cite{Schrodinger1930} and includes also the
anticommutator $\{A,B\}$, namely
\begin{eqnarray}
\lefteqn{(\Delta A)^2 (\Delta B)^2 \geq } \nonumber \\
&& \left( \frac{1}{2} \left\langle\{A, B\} \right\rangle- \langle A \rangle \langle B \rangle \right)^2 + \left( \frac{1}{2i} \left\langle[A,B] \right\rangle \right)^2
\label{SR}
\end{eqnarray}
with $(\Delta O)^2=\langle {O}^2 \rangle -
\langle O \rangle^2$ being the variance of observable~$O$. If
one does not consider the first squared term in the right-hand side (rhs) of~\eqref{SR}, one deals with the
usual Heisenberg--Robertson~(HR) uncertainty relation.

In the particular case of observables~\eqref{A}
and~\eqref{B}, the
Schr\"odinger--Robertson~(SR) uncertainty relation~\eqref{SR} reads
\begin{eqnarray} \label{SRspin}
\lefteqn{\left[1- (\vec a \cdot \vec s)^2\right] \left[1- (\vec b \cdot \vec s)^2\right] \geq} \nonumber
\\ && \left[ \vec a \cdot \vec b- (\vec a \cdot \vec s)(\vec b \cdot \vec s) \right]^2+ \left[ (\vec a \times \vec b) \cdot \vec s \right]^2
\end{eqnarray}
where we used that \ $\langle A \rangle= \alpha_1+ \alpha_2\, \vec a \cdot
\vec s$, \ $(\Delta A)^2 =\alpha_2^2 \left[1- (\vec a \cdot \vec s)^2 \right]$, and
analogously for $B$, while
$\{A,B\}=2\left[ \left( \alpha_1 \beta_1 + \alpha_2 \beta_2 \, \vec a \cdot \vec b\right) I + \alpha_2 \beta_1 \, \vec a \cdot \vec \sigma + \alpha_1 \beta_2 \, \vec b \cdot \vec \sigma \right]$ \ and \ $[A,B]=2i \alpha_2 \beta_2 (\vec a \times \vec b)\cdot\vec\sigma$.

Variance-based UP formulations have been doubly criticized. On the
one hand, the lower bound to the product of variances depends, in
general, on the state of the system via the expectation values and
thus lacks a universal character~\cite{Deutsch1983,Maassen1988}.
Moreover, it can be easily seen~\cite{Ghirardi2003} that for
discrete, bounded operators the lower bound is trivially zero,
yielding no valuable information. On the other hand, the use of the
variance as measure of uncertainty (spreading) of a given
probability distribution exhibits some
limitations~\cite{Hilgevoord2002,Birula2011}.
It might also be the case that  the variance is not well-defined.

\subsection{Landau-Pollak uncertainty relation}

An alternative UP formulation was introduced by Landau and
Pollak (LP) in the context of time-frequency analysis~\cite{Landau1961}, and adapted to the quantum framework by Maassen
and Uffink~\cite{Maassen1988}. Using the notation $M_{\infty}(A;
\rho) = \max_i p_i(A)$ for the maximum probability of the outcomes
of observable $A$, then the LP uncertainty relation reads
\begin{eqnarray}\label{LP}
&\arccos \sqrt {M_{\infty}(A; |\Psi\rangle
\langle\Psi|)}&\nonumber\\
&+ \arccos \sqrt {M_{\infty}(B; |\Psi\rangle \langle\Psi|)} \geq
\arccos c&
\end{eqnarray}
The LP relation captures the essence of the uncertainty principle
for quantum pure state, indeed the rhs is
state-independent.
A generalization of a weak version of the LP inequality for positive
operator valued measures was recently given in
Ref.~\cite{Miyadera2007}, and an application of this inequality to
separability problems was developed  in Ref.~\cite{deVicente2005}.

For $N$-dimensional systems, the extension of~\eqref{LP} to general (mixed) states is not obvious, due to the lack
of definite concavity of $\arccos \sqrt {M_{\infty}(A; \rho)}$.
However, for two-dimensional systems it can been shown that the LP
relation remains valid for mixed states.

For our purposes, we express the LP inequality~\eqref{LP} as
\begin{equation}\label{LP2}
\sqrt{M_{\infty}(A) M_{\infty}(B)}-\sqrt{[1- M_{\infty}(A)]
[1-M_{\infty}(B)]} \leq c
\end{equation}
where we have used the trigonometric identity \ $\arccos x
+ \arccos y = \arccos \left( xy-\sqrt{(1-x^2)(1-y^2)} \right)$ for
$x+y \geq 0$ \ \cite{Table}, and that $\arccos(x)$ is a decreasing
function. In our 2D case, using~\eqref{pA}--\eqref{pB}, we have
\begin{eqnarray}
\sqrt{(1+|\vec{a}\cdot \vec{s}|) (1+|\vec{b}\cdot \vec{s}|)} - \sqrt{(1-|\vec{a}\cdot \vec{s}|) (1-|\vec{b}\cdot \vec{s}|)} && \nonumber \\
\leq 1+|\vec{a}\cdot\vec{b}| &&
\end{eqnarray}

\subsection{Entropy-based uncertainty relations}

Information-theory tools have shown their usefulness in the study
of uncertainty relations~\cite{Mamojka1974,Birula1975}. Consider
now, as a measure of uncertainty (ignorance), the one parameter
generalization of Shannon entropy given by
R\'enyi~\cite{Renyi1970}, that in the case of an $N$-dimensional,
discrete probability distribution reads
\begin{equation}\label{Renyi}
H_q (\{p_i\})= \frac{1}{1-q} \ln \left( \sum_{i=1}^N p_i^{\ q} \right)
\end{equation}
where $0 \leq p_i \leq 1$, \ $\sum_{i=1}^N p_i=1$, and the real parameter $q>0$ with $q\neq
1$. If we let $q\rightarrow 1$, then this definition includes by continuity the
Shannon case: \ $H_1(\{p_i\})=- \sum_{i=1}^N p_i \ln p_i$. Other
special $q$ values of interest, for instance in quantum information
process and quantum cryptography, are $q=2$ and $q\rightarrow
\infty$. In the former case, \ $H_2(\{p_i\})= - \sum p_i^{\,2}$ is
known as collision entropy. The latter is known as min-entropy, due
to the property $H_{q'}<H_q$ if $q'>q$ for fixed $\{p_i\}$, and its
value is \ $H_\infty(\{p_i\})=-\ln(\max_i\{p_i\})$.

An \emph{entropic uncertainty relation} (EUR) has the form
\begin{equation}\label{e:EUR}
H(A;\rho)+ H(B;\rho) \geq \mathcal{B}(A,B)
\end{equation}
where $H$ is an entropic measure like the ones in
Eq.~\eqref{Renyi} with the probability distributions of the
observables calculated via Born rule, while $\mathcal{B}$ is a
function of them. More precisely, it depends on the overlap between
both eigenbases, being state-independent (i.e.\ it is not a function
of the state $\rho$) and a positive quantity. The search of
tight bounds $\mathcal{B}$ for different pairs of observables with
discrete or continuous spectra, using diverse entropic forms, has
been subject of intense interest, for instance in Refs.~\cite{Zozor2008,deVicente2008,Wehner2010,Birula2011}.

In the particular case of spin-$1/2$ observables, using Eq.~\eqref{pA}, the R\'enyi entropy reads
\begin{equation} \label{Hq}
H_q ( A;\rho) = \frac{1}{1-q} \ln \left[ \left( \frac{1+ \vec{a} \cdot \vec{s} }{2} \right)^q + \left( \frac{1- \vec{a} \cdot \vec{s} }{2} \right)^q \right]
\end{equation}
Note that this is a measure of the degree of uncertainty
associated to the observable $A$, in the following sense: when one
is certain about the observable's value, i.e.\
$\{p(A)\}=\{1,0\}$ or $\{0,1\}$, then the entropy takes its
minimum value $H_q=0$. Contrariwise, for total ignorance
concerning the value of $A$, i.e.\
$\{p(A)\}=\{\frac{1}{2},\frac{1}{2}\}$, the entropy is
maximal and equal to $H_q=\ln 2$ (irrespective of $q$). R\'enyi
entropy is a concave function in $\rho$ for $q$ lying in the
interval $(0,2]$, that is, if $\rho = \sum_n \lambda_n
|\Psi_n\rangle \langle\Psi_n|$, with $0 \leq \lambda_n \leq 1$ and
$\sum_n \lambda_n = 1$, then $H_q ( A;\rho) \geq \sum_n \lambda_n
H_q (A;|\Psi_n\rangle \langle\Psi_n|)$ \
\cite{Ben1978,Bengtsson2006}. In the following we restrict the
value of $q$ to the above interval.

Optimal entropic uncertainty relations for two arbitrary quantum
observables in the 2-dimensional case were
obtained for the Shannon~\cite{Ghirardi2003} and collision~\cite{Bosyk2012} entropies. Specifically, when $A$ and $B$ are spin-1/2
complementary observables ($c=1/\sqrt 2$) the optimal lower bounds are $\ln 2$ and
$2 \ln 4/3$, respectively.

\section{Connections between complementarity and uncertainty relations}
\label{s:Relationship}

\subsection{Equivalence with variance-based uncertainty relations}

The relationship between the predictability--visibility
inequality~\eqref{CP} and the uncertainty relations based on
variances~\eqref{SR} are readily analyzed using the Bloch
representation of the pertinent operators and the
density matrix. First of all, the variances of the operators
defined in Eqs.~\eqref{Pop}--\eqref{Vop2} are given, in terms of
the predictability~$P$ and visibility~$V$, by
\begin{eqnarray}
(\Delta \hat{P})^2&=& 1- P^2 \label{DP}\\
(\Delta \hat{V}_{\phi})^2&=& 1- V^2 \cos^2(\theta -\phi) \label{DV} \\
(\Delta \hat{V}_{\phi}^{\perp})^2&=& 1- V^2 \sin^2(\theta -\phi) \label{DVO}
\end{eqnarray}
The connection between the CP relation and variance-based URs has
been analyzed in Refs.~\cite{Durr2000}, \cite{Bjork1999},
and~\cite{Busch2006}. In~\cite{Durr2000} the authors highlight the
{\it equivalence} between both  principles. Indeed, they compute
the HR-UR for the pair of observables $\hat{P}$ and
$\hat{V}_\theta^\perp$, and also for $\hat{V}_\theta$ and
$\hat{V}^\perp_\theta$ (setting the phase shifter  to an angle
$\phi=\theta$). By doing so, they obtain the following uncertainty
inequalities
\begin{eqnarray}
(\Delta \hat{P})^2 (\Delta \hat{V}_\theta^\perp)^2 & = &  1-P^2 \geq V^2 \\
(\Delta \hat{V}_\theta)^2 (\Delta \hat{V}_\theta^\perp)^2 & = &
1-V^2 \geq P^2
\end{eqnarray}
and notice that both are equivalent
to~\eqref{CP}. The main drawback that they note in their
derivation is the use of $\hat{V}^\perp_\theta$, which has no direct
interpretation in terms of neither predictability nor visibility in
connection with the MZ interferometry experiment, since
$\langle\hat{V}^\perp_\theta\rangle=0$ and $\Delta\hat{V}^\perp_\theta=1$. Moreover, when dealing with
$\hat P$ and $\hat{V}_\theta$, the corresponding HR-UR becomes
trivial: \ $(\Delta \hat{P})^2 (\Delta \hat{V}_\theta)^2 \geq 0$.

Independently, Bj\"ork {\it et al.} also dealt with the problem of
connecting CP with UP. Although the authors of
Ref.~\cite{Bjork1999} mention the SR-UR, they do not actually
calculate the lower bound of the product of variances as
prescribed by the rhs of~\eqref{SR}. Their analysis is, in this
respect,  limited to obtain expressions~\eqref{DP}
and~\eqref{DV} followed by an appeal to $(\Delta \hat{V}_{\phi})^2
\geq (\Delta \hat{V}_{\theta})^2$ (basic trigonometry) with the
purpose of linking the two fundamental principles of quantum
mechanics.

A complete proof of the alluded to equivalence dealing with the {\it
appropriate} observables $\hat P$ and $\hat{V}_\theta$ and the
{\it full} SR-UR, is given in Ref.~\cite{Busch2006}. We
reproduce it here --although in a slightly different way-- for the
sake of completeness. For arbitrary $\phi$, the UR prescribed by Schr\"odinger and Robertson reads
\begin{eqnarray}\label{SR-PV}
\lefteqn{ (1- P^2) [1-V^2\cos^2(\theta-\phi)] \geq } \nonumber \\
&& P^2 V^2 \cos^2(\theta-\phi)+V^2 \sin^2(\theta-\phi)
\end{eqnarray}
where equality holds for any pure state. It is straightforward to
show that this inequality is \emph{equivalent} to the duality
relation~\eqref{CP}. We stress that~\eqref{SR-PV} is valid for
{\it any} phase $\phi$ introduced by the phase shifter in the MZ
interferometer. We then conclude that the appropriate choice
$\phi=\theta$ implies \emph{equivalence} with the trade-off
relation between predictability and visibility. This circumvents
the drawback pointed out by D\"urr and Rempe. With this simple
result, a rather sharp conclusion is drawn from the discussion
about complementarity between $P$ and $V$, including the status
of~\eqref{CP} as an uncertainty relation. Finally, we mention that
in Ref.~\cite{Liu2012} a relation between wave--particle duality
and quantum uncertainty has been investigated, both theoretically
and experimentally, by recourse to variances of the operators
$\hat P$ and $\hat V_{\theta}$, although without appealing to
Heisenberg-like inequalities.

\subsection{Equivalence with Landau--Pollak uncertainty relation}

Let us now see just how inequality~\eqref{CP}
becomes equivalent to Landau--Pollak uncertainty relation.
The maximum probabilities associated to observables $\hat P$ and $\hat
{V}_\theta$, in terms of the predictability and visibility, are
\begin{eqnarray}
M_\infty(\hat P)&=& \frac{1+P}{2} \\
M_\infty (\hat V_\theta) &=& \frac{1+V}{2}
\end{eqnarray}
Replacing these probabilities in~\eqref{LP2}, and setting
$c=1/\sqrt2$ as corresponds to the case of complementary operators, we obtain
\begin{equation}
\sqrt{ \left( \frac{1+P}{2} \right) \left( \frac{1+V}{2} \right)} -
\sqrt{ \left( \frac{1-P}{2} \right) \left( \frac{1-V}{2} \right)}
\leq \frac{1}{\sqrt 2}
\label{LP-PV}
\end{equation}
Squaring both sides of this inequality and grouping terms
conveniently, we immediately arrive at the relation
\begin{equation}
(1-P^2)(1-V^2) \geq (P V)^2
\end{equation}
which coincides with~\eqref{SR-PV} for $\phi=\theta$, and, as mentioned before, can be easily recast in the fashion \
$P^2+V^2 \leq 1$. This implies that the duality relation~\eqref{CP} can
be deduced from the LP inequality, and viceversa.

\subsection{Relationship with entropic uncertainty relations}

Having clarified the above equivalences, we now consider the problem of
elucidating the connection between EURs and the duality relation~\eqref{CP}. Using Eq.~\eqref{Hq} and the Bloch representation of
$\hat P$ and $\hat{V}_\theta$ we obtain
\begin{eqnarray}\label{e:SumOfEntropies}
H_q(P)= \frac{1}{1-q} \ln \left[ \left( \frac{1+ P}{2} \right)^q + \left( \frac{1- P}{2} \right)^q \right] \\
H_q(V)=\frac{1}{1-q} \ln \left[ \left( \frac{1+
V}{2} \right)^q + \left( \frac{1- V}{2} \right)^q \right]
\end{eqnarray}
where to simplify notation we have renamed $H_q(\hat P;\rho) \equiv H_q(P)$ and $H_q(\hat{V}_\theta;\rho)\equiv H_q(V)$.
Our goal is to find the minimum of the sum of these R\'enyi
entropies over all available states, that is, $\displaystyle \min_{\rho} \{H_q(\hat P;\rho)+ H_q(\hat{V}_\theta;\rho)\}$.
Appealing to the concavity of R\'enyi entropy for $q\in(0,2]$, we can restrict
our calculations to pure states and then the conditioned
minimization problem can be recast in the fashion
\begin{equation}\label{e:min2}
\min_{P^2+V^2=1} \{H_q(P)+H_q(V) \}
\end{equation}
For arbitrary values of $q$, this problem can be solved numerically. 
%(for a detailed calculation see~\cite{Zozor2012}). %%
It is seen that three qualitatively different regimes appear: (i)~for $0 < q < q^*$ with $q^*\approx 1.4316$,
the minimum is $\ln 2$  and it is attained at $V=0$ and $P=1$, or $V=1$ and $P=0$;
 (ii)~at $q=q^*$ the minimum value is also $\ln 2$ but it corresponds to the cases $V=0$ and  $P=1$, $V=1$ and $P=0$, or also $V=P=1/\sqrt2$; and (iii)~for $q^* < q \leq 2$, the minimum is the $q$-dependent function \
$\frac{2}{1-q} \ln \left[ \left( \frac{1+ 1/\sqrt2}{2} \right)^q +
\left( \frac{1- 1/\sqrt2}{2} \right)^q \right]$, attained at $V=P=1/\sqrt2$. The value of $q^*$ is obtained solving (numerically) the equation \ $2H_{q^*}(1/\sqrt2)=\ln 2$.

\begin{figure}[htbp]
\begin{center}
\includegraphics[width=8cm]{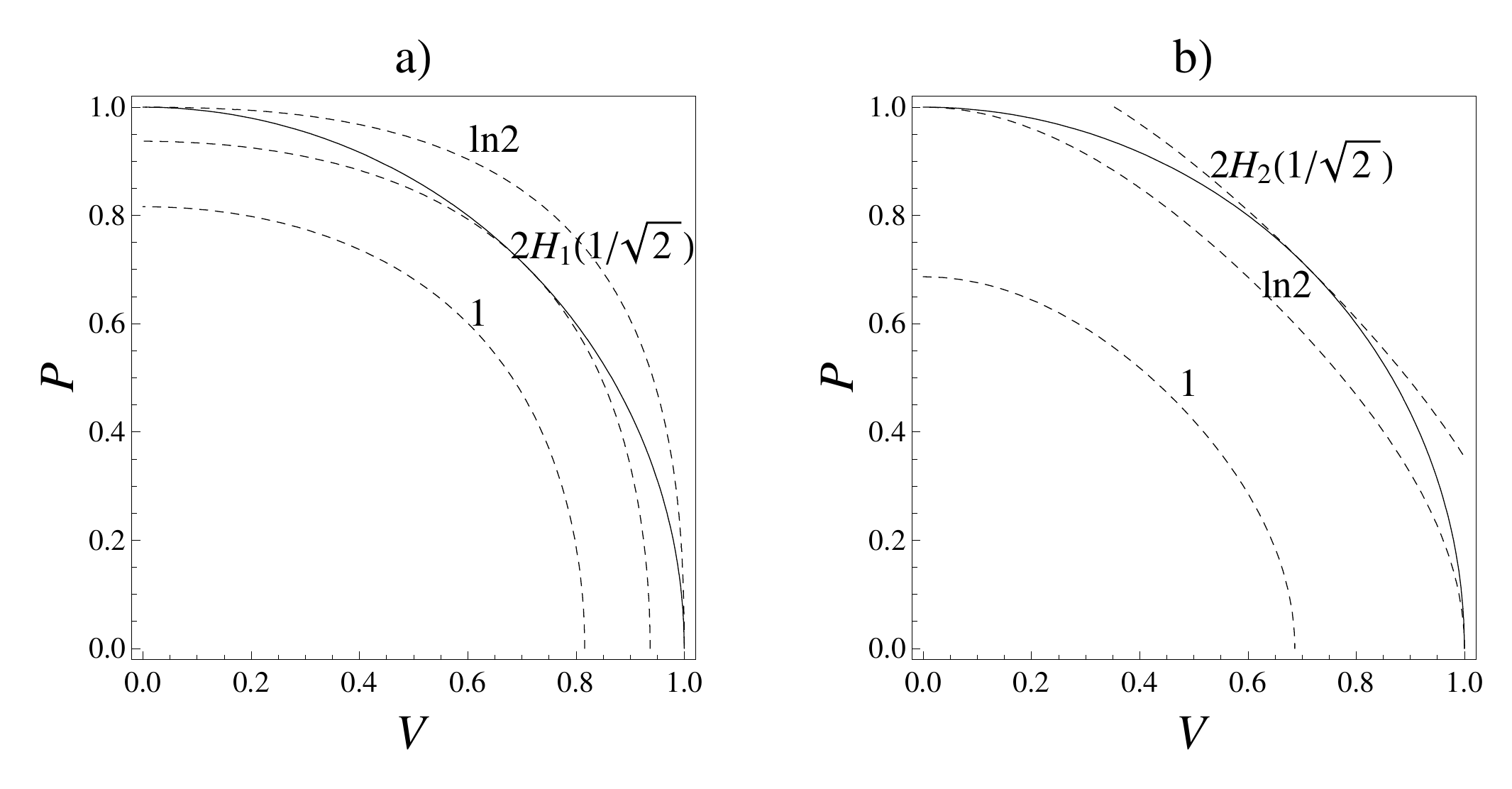}
\end{center}
\caption{Constraint $P^2+V^2=1$ (solid line) and contour plots (dashed lines) of the sum of $q$-R\'enyi entropies with entropic indices: a) $q=1$, b) $q=2$. The value of the entropy sum is indicated near each contour line: the values shown are 1, $2H_1(1/\sqrt2) \approx 0.833$, $\ln 2 \approx 0.693$, and $2H_2(1/\sqrt2) \approx 0.576$. }
\label{f:figure2}
\end{figure}

In Fig.~\ref{f:figure2} we display, in the $V$-$P$ plane, the constraint $P^2+V^2=1$ together with several contour lines of the sum of $q$-R\'enyi entropies for two representative values of the entropic parameter in the regimes~(i) and~(iii) mentioned above. In both cases
the contour lines correspond to decreasing values towards the origin. In case (i) $\ln 2$ is the minimum-value contour line that intersects (tangencially) the constraint, at the points $(V,P)=(0,1)$ or $(V,P)=(1,0)$. In case (iii) the curve $P^2+V^2=1$ is intersected by the minimum-value contour line $H_q(P)+H_q(V)=2H_q(1/\sqrt 2)$, precisely at $(V,P)=(1/\sqrt2,1/\sqrt2)$.

The existence of these three regimes sheds light on the
meaning of the sum of R\'enyi entropies. The fact that there appear three
qualitatively different regimes agrees with a result
suggested in previous work~\cite{Luis2011}, where different values
of the parameter $q$ yield different entropic measures which, in turn, give rise to qualitatively different information about the system.
As stated before, information-theoretic entropy gives a measure of the uncertainty related to the outcome of a variable in terms of the corresponding probability
distribution. Thus, solving the problem raised in~\eqref{e:min2} we get the
minimum of the sum of uncertainties (notice that in this spirit we are using the same $q$ parameter for both entropies). We are also able to pinpoint the optimum (minimizing) states of this problem.

States which saturate an uncertainty relation are used in several
contexts. An important example has to do with coherent states, which
saturate the position--momentum Heisenberg UR. Because of
having this property, coherent states are usually interpreted as the
most classical ones. In the present case, in which we consider the predictability--visibility relation in the context of the MZ interferometer, we find that regime (i) is a trivial one, being the eigenstates of $\hat V_{\theta}$ or $\hat P$ those states of minimum uncertainty sum. However, an interesting non-trivial situation
appears in regimes~(ii) and~(iii), where the extremum is
attained at the symmetric case.
Which are the characteristic features of states which make
$|\langle \hat V_{\theta}\rangle|=|\langle \hat P\rangle|=\frac{1}{\sqrt{2}}$? They are the pure states of the form \eqref{e:stateBloch} with the four different unit Bloch vectors: \  $\pm\left(\frac{1}{\sqrt 2}\cos\theta,\frac{1}{\sqrt 2}\sin\theta,\pm\frac{1}{\sqrt 2}\right)$.
These are precisely the states
which saturate the concomitant EURs in the most
unbiased way (in the sense of simultaneously having the maximum visibility {\it and} maximum predictability that is possible).

Let us consider in more detail the question of getting the states with maximum value for both predictability and visibility, a situation which one expects to correspond to the best description of the system. From simple geometric arguments in the $V$-$P$ plane (taking into account that $P^2+V^2\leq 1$ has to be fulfilled), it is
straightforward how to compute those states. However, one may deal as well with a situation in which one does not have at hand
the whole set of states available, but only a fraction of it. In such circumstances, it is convenient to delve further into the usefulness of the minimization of the measure $H_q(P)+H_q(V)$ (when $q>q^*$). For
example, this situation may appear if the source in Fig.~\ref{f:figure1} has limitations for producing certain states, and one is thus
restricted to deal with a given region of the convex set of quantum states.
Another interesting situation has to do with the case in which the
second beam splitter is a Schr\"odinger cat (as is the case in
Refs.~\cite{Auccaise-2012} and~\cite{Roy-2012}) or if there is a noisy
environment. In both situations, the states of the system which pass trough
the interferometer are limited by the state of the environment
(and cannot be controlled, in the second case), being mixed states
the more general case. Thus, not all states are available and~\eqref{e:min2} gives a way to solve the
problem posed by conditions mentioned above in this non-symmetrical
setting.

\section{Conclusions}\label{s:Conclusions}

We studied here connections between the Complementarity
and Uncertainty Principles in the Mach--Zehnder interferometer
scheme. Following Ref.~\cite{Bjork1999} and related work, we employed quantum-mechanical operators $\hat P$
and $\hat V_{\theta}$ to represent the particle and wave aspects of a quantum system,
respectively.

We have thoroughly analyzed some drawbacks concerning the approaches of Bj\"ork {\it et al.}, and of D\"urr and
Rempe, who considered the link between the inequality~\eqref{CP} and variance-based uncertainty relations of the form~\eqref{SR}.
We showed the {\it equivalence} between the Schr\"odinger--Robertson UR and the duality relation in the relevant case, i.e.\ for observables which adequately represent predictability and visibility according to~\cite{Busch2006}.
An alternative quantification of the Uncertainty Principle is given by the Landau--Pollak inequality~\eqref{LP}. We proved the equivalence between~\eqref{CP} and the LP-UR (as specified for the observables of interest).

It is worth stressing then that in the present context the three inequalities~\eqref{CP},~\eqref{SR-PV} and~\eqref{LP-PV}, are on an equal footing (which may well not be the case for other pairs of observables).
We remark that our
results give a precise (and quantitative) meaning to the assertion
\emph{$P$ and $V$ are complementary quantities} and, at the same
time, settle pending question regarding the status
of~\eqref{CP} as an uncertainty relation.

Moreover, we have studied the connection between~\eqref{CP} and
entropic uncertainty relations~\eqref{e:EUR} based on the $q$-R\'{e}nyi entropy~\eqref{Renyi}. We found that these EURs, for the pair $\hat P$-$\hat V_{\theta}$, are not equivalent to the duality relation. Nevertheless, we see that, when these uncertainty measures are applied
to the MZ scheme, different regimes emerge, depending on the value
of the entropic parameter~$q$.
We also noticed that this agrees with a previous investigation by Luis~\cite{Luis2011}, in which the value chosen for $q$ affects the qualitative behavior of the uncertainty relations.
The fact
that these different regimes are also found in the canonical example
of MZ interferometry seems to provide support to the assertion
{\it there is no preferred value for $q$}. %% in this context
Indeed, different $q$-values render the concomitant entropic measures useful for different purposes.
In addition, looking at the states which correspond to an equality in the
entropic uncertainty relation, we find regimes with nontrivial saturating states. We have in this vein established a procedure for solving the problem of finding a state having minimum
uncertainty for the observables $\hat P$ and $\hat V_{\theta}$ in the most unbiased fashion. Finally, we
also discussed the usefulness of such procedure for infomation-theoretical purposes, depending on the nature of the source and the
beam splitters.

\section*{ACKNOWLEDGMENTS}

This work has been supported by PICT-2007-806 (ANPCyT) and PIP~1177/09 (CONICET), Argentina.

\end{document}